\documentclass[12pt,twoside,fleqn]{article}
\usepackage{epsfig}
\usepackage{psfig}
\def\lsim{\raise0.3ex\hbox{$<$\kern-0.75em\raise-1.1ex\hbox{$\sim$}}}
\def\gsim{\raise0.3ex\hbox{$>$\kern-0.75em\raise-1.1ex\hbox{$\sim$}}}
\newcommand{\beqn} {\begin{equation}}
\newcommand{\eqn} {\end{equation}}

\newcommand{\plaq}{\mbox{\raisebox{-.75mm}
{\epsfig{file=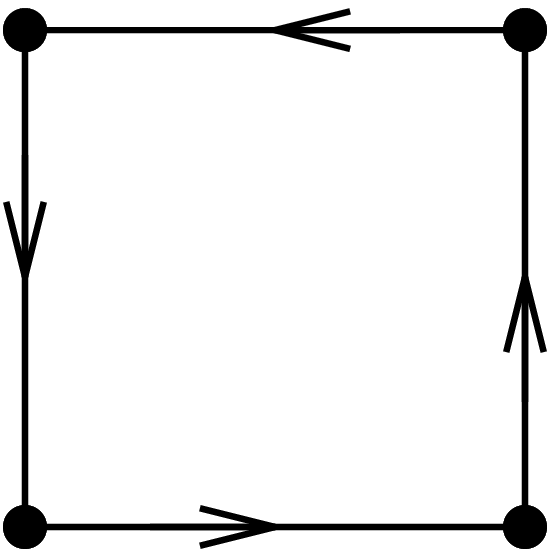,height=3mm
}}~}}

\newcommand{\loOp}{\mbox{\raisebox{-.75mm}
{\epsfig{file=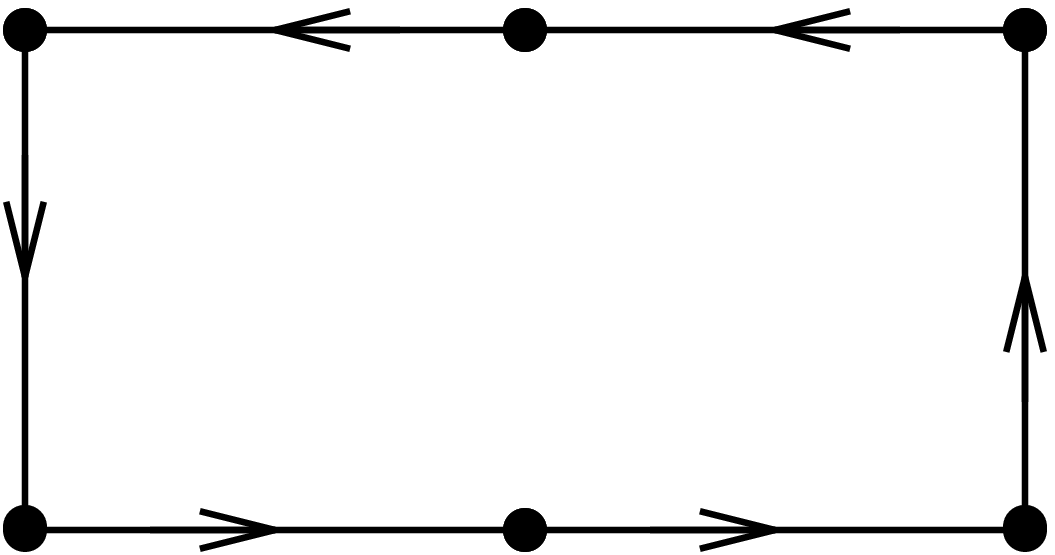,height=3mm
}}~}}
\newcommand{\lOop}{\mbox{\raisebox{-2mm}
{\epsfig{file=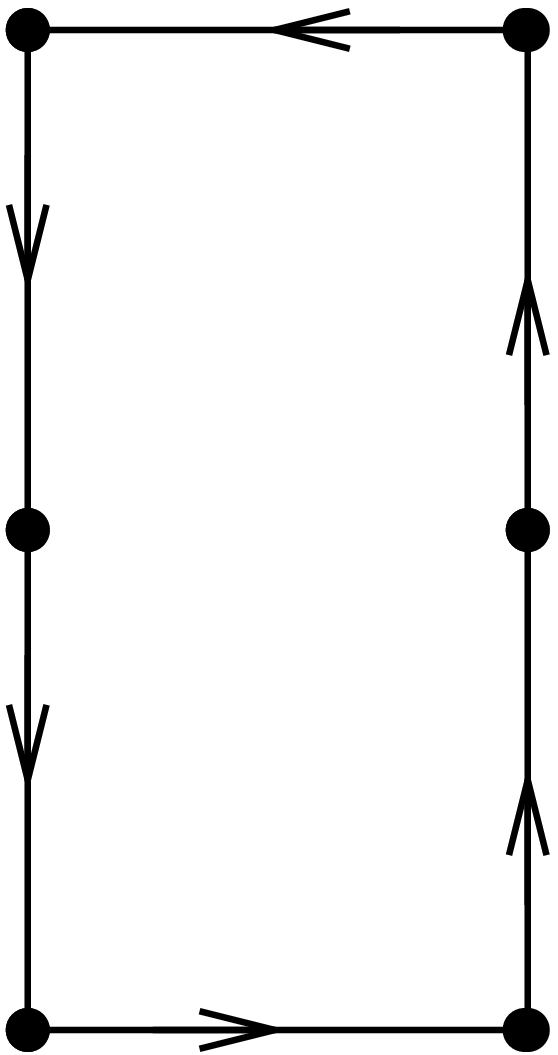,height=6mm
}}~}}
\newcommand{\nn}{\nonumber}
\newcommand{\tr}{\mbox{Tr~}}
\newcommand{\re}{\mbox{Re~}}
\newcommand{\kronecker}[2]{  \delta_{#1 , #2}  }

\newcommand{\mymatrix}[2]{
        m \kronecker{#1}{#2} + \eta_{#1}\left( \;
                    \frac{9}{16} A\left[ U \right]_{#1\; #2} -
        \frac{1}{48} B\left[ U  \right]_{#1\; #2} \right) }
\newcommand{\linkdagger}[2]{  U_{ #1 , #2}^\dagger  }
\newcommand{\link}[2]{  U_{ #1 , #2}  }
\newcommand{\apart}[3]{
\sum_{#3} \left( \link{#1}{#3} \kronecker{#1}{#2-\hat{#3}}
                -\linkdagger{#1-\hat{#3}}{#3}
                                        \kronecker{#1}{#2+\hat{#3}} \right) }

\newcommand{\bparta}[3]{
\sum_{#3} ( \link{#1}{#3} \link{#1+\hat{#3}}{#3} \link{#1+2 \hat{#3}}{#3}
                                             \kronecker{#1}{#2- 3 \hat{#3}}   }
\newcommand{\bpartb}[3]{
                 \linkdagger{#1-\hat{#3}}{#3} \linkdagger{#1-2\hat{#3}}{#3}
                 \linkdagger{#1-3 \hat{#3}}{#3} \kronecker{#1}{#2+ 3 \hat{#3}}
                  ) }

\newcommand{\AmS}{{\protect\the\textfont2
  A\kern-.1667em\lower.5ex\hbox{M}\kern-.125emS}}

\begin{document}
%
\mbox{} \hfill BI-TP 96/58\\
\mbox{} \hfill December 1996\\
\begin{center}
{{\large \bf Thermodynamics of Four-Flavour QCD \\
with Improved Staggered Fermions}
} \\
\vspace*{1.0cm}
{\large  J. Engels, R. Joswig, F. Karsch, E. Laermann, \\
M. L\"utgemeier and B. Petersson }\\
\vspace*{1.0cm}
{\normalsize
$\mbox{}$ {Fakult\"at f\"ur Physik, Universit\"at Bielefeld,
D-33615 Bielefeld, Germany}
}
\end{center}
\vspace*{1.0cm}
\centerline{\large ABSTRACT}

\baselineskip 10pt

\noindent
We have calculated the pressure and energy density in four-flavour 
QCD using improved fermion and gauge actions. We observe a
strong reduction of
finite cut-off effects in the high temperature regime, similar to what
has been noted before for the SU(3) gauge theory. Calculations have been
performed on $16^3\times 4$ and $16^4$ lattices for two values of the
quark mass, $ma = 0.05$ and 0.1. A calculation of the string tension
at zero temperature yields a critical temperature
$T_c/\sqrt{\sigma} = 0.407 \pm 0.010$ for the smaller quark mass value.

\vskip 20pt
\vfill
\eject
\baselineskip 15pt
\noindent
\section{Introduction}
\vskip 5pt

\noindent
In the pure gauge sector of QCD improved actions have been shown to
lead to a drastic reduction of systematic errors introduced by
the non-zero lattice spacing ({\it finite cut-off effects}) in the
calculation of thermodynamic observables \cite{Bei96,Pap96,Bei96a}.
This is particularly evident
in the high temperature regime where analytic calculations in the
infinite temperature, ideal gas limit show that deviations from the
continuum result can be drastically reduced already with tree level
improved Symanzik actions \cite{Bei96}. However, also close to $T_c$ and
even at $T_c$ an improvement has been observed already with tree level
improved actions \cite{Bei96a}.

At non-zero temperature the finite cut-off effects become visible as a
{\it finite size effect} because the lattice spacing in units of the
temperature is given by the temporal extent, $N_\tau$, of the lattice,
{\it i.e.}  $aT= 1/N_\tau$. In the standard Wilson formulation of
lattice QCD with staggered fermions these cut-off effects are known to
be ${\cal O} ((aT)^2 \equiv 1/N_\tau^2)$. In the
high temperature, ideal gas limit one finds, for instance, for the
gluonic and fermionic contributions to the energy density of QCD with $n_f$
massless flavours
\begin{eqnarray}
{\epsilon^G \over T^4} &=& {8\pi^2 \over 15} \biggl(1 + {30\over 63}
{\pi^2\over N_\tau^2} + {\cal O} (N_\tau^{-4}) \biggr)~~, \nn \\
{\epsilon^F \over T^4} &=& n_f{21\pi^2 \over 60} \biggl(1 + {310\over 147}
{\pi^2 \over N_\tau^2} + {\cal O} (N_\tau^{-4}) \biggr)~~.
\label{freee}
\end{eqnarray}
On lattices
with small temporal extent these ${\cal O} (N_\tau^{-2})$ corrections 
lead to large distortions of
the continuum Stefan-Boltzmann limit \cite{Kar82}. In the
staggered fermion
formulation the deviations from the continuum ideal gas are as large as
77\% on a lattice with temporal extent $N_\tau=4$ and are still
20 \% for $N_\tau= 12$.

As the computational effort for calculating
thermodynamic observables, which generally have dimension $a^{-4}$,
increases approximately like $N_\tau^{10}$ it is highly desired to be able to
extract continuum physics from results of
simulations on lattices with small temporal extent. In addition 
the improvement of the fermion action rapidly becomes
computationally very demanding. In the inversion of the
fermion matrix the time per iteration is proportional to the number of non-zero
entries in a given row of the fermion matrix, {\it i.e.} the number of
neighbours to a given lattice site used to discretize the
kinetic part of the Lagrangian. In particular in simulations with
dynamical fermions this gives a very serious constraint and does, at
present, make it impossible to use, for instance, a complicated fixed point
action \cite{Bie96}. In this
first exploratory study, which aims at an examination of the
influence of an improved fermion sector on the behaviour of
thermodynamic observables at high temperature, we therefore have chosen
a straightforward improvement scheme for the free fermion sector.
The minimal extension of the standard staggered fermion discretization
scheme which is capable to remove ${\cal O}(a^2)$ errors in the free
fermion action involves in
addition to the standard one-link term an appropriately weighted,
straight three-link term.
This preserves all the symmetries of the staggered action
\cite{Nai89} and doubles the computational effort in the fermion
sector.

In the next section we specify the improved action used here and
discuss the cut-off dependence of thermodynamic observables in the
infinite temperature, ideal gas limit. Section 3 is devoted to a
discussion of the equation of state of four-flavour QCD. In section 4 we
discuss a calculation of the critical temperature in units of the string
tension. Finally we present our conclusions in section 5.

\section{Improved actions}
\vskip 5pt

\noindent
When formulating a discretized version of QCD one has a great deal of
freedom in choosing a lattice action. Different formulations may differ by
subleading powers of the lattice cut-off, which vanish in the continuum
limit. This has, for instance, been used to systematically
improve lattice regularized $SU(N)$ gauge theories
following the Symanzik improvement programme \cite{Sym83,Wei83}. In
addition to the elementary
plaquette term appearing in the standard Wilson formulation of lattice QCD
larger loops can be added to the action in such a way that the leading
${\cal O} (a^2)$ deviations are eliminated from the free fermion action. 
In our previous studies of
the pure gauge sector of QCD \cite{Bei96,Bei96a} we have found that
already the addition of a planar (1,2) loop to the standard Wilson
action eliminates the main systematic errors in the high temperature
limit and even at $T_c$ compares well with a non-perturbatively improved
(tadpole) action \cite{Lep93} and a fixed point action \cite{Pap96}.
We therefore use the tree-level improved (1,2)-Symanzik action for
discretizing the pure gauge sector of QCD,

\begin{eqnarray}
\hskip -2.0truecm S^{(1,2)}\hskip -0.3cm &=& \hskip -0.3cm
\sum_{x, \nu > \mu}~ {5\over 3}~\left(
1-\frac{1}{N}\re\tr\plaq_{\mu\nu}(x)\right)
\nn\\
& &\hskip 0.3truecm
-{1\over 6 }\left(1-\frac{1}{2N}\re\tr
\left(\loOp_{\mu\nu}(x)+\lOop_{\mu\nu}(x)\right)\right)~~.
\label{actiong}
\end{eqnarray}
In the fermion sector we use an improved action, $S^F_{3}=
\bar{\psi} M \psi$, where a higher order
difference scheme is used to eliminate
the ${\cal O} (a^2)$ errors in the discretization of the derivatives
$\partial_\mu$ appearing in the free fermion action. A
three-link term is added which preserves all the symmetries of the
staggered action \cite{Nai89}. Gauge fields are then introduced on the
shortest path connecting the staggered fermion and anti-fermion fields.
The improved fermion matrix thus reads
\beqn
M [U]_{ij}\hskip -0.2cm = \mymatrix{i}{j}
\label{fermionM}
\eqn
with $\eta_i$ denoting the phase factors for staggered fermions and
\begin{eqnarray}
A[U]_{ij}\hskip -0.2cm
&=&\hskip -0.1cm \apart{i}{j}{\mu} \nn \\
B[U]_{ij}\hskip -0.2cm &=&\hskip -0.2cm
\bparta{i}{j}{\mu} -
\hskip -0.1cm   \bpartb{i}{j}{\mu} ~~.
\label{fermionAB}
\end{eqnarray}
The entire action is given by  $S^I[U] = \beta S^{(1,2)} + S^F_{3}$.

The importance of improved actions for thermodynamic calculations 
becomes evident
from an analysis of the high temperature ideal gas limit on lattices of
size $N_\sigma^3  N_\tau$.  One indeed finds a strong reduction
of the cut-off dependence relative to the standard Wilson formulation.
In order to quantify this we previously had calculated
the energy density of a free gluon gas \cite{Bei96} which shows
explicitly that corrections to the continuum Stefan-Boltzmann law only
start at ${\cal O} (N_\tau^{-4})$. We give here the corresponding results 
for a free massless fermion gas using the improved fermion
action for calculations on spatially infinite lattices with
temporal extent $N_\tau$,

\begin{eqnarray}
{\epsilon^F (N_\tau) \over T^4} &=& 3n_f N_\tau^4 \int_0^{1} {\rm d}^3\vec{p}
\biggl[ N_\tau^{-1} \sum_{n_0=0}^{N_\tau-1} {f^2\bigl((2n_0+1)\pi/N_\tau\bigr)
\over \omega^2(2\pi\vec{p}) +4 f^2\bigl((2n_0+1)\pi/N_\tau\bigr) } \nn \\
& & - \int_0^{1}{\rm d}p_0 {f^2(2\pi p_0)
\over \omega^2(2\pi\vec{p}) +4 f^2(2\pi p_0) }\biggr] ~~,
\label{energyf}
\end{eqnarray}
where $n_0=0,$ 1,..., $N_\tau-1$ labels the discrete set of Matsubara
modes, $\omega^2(2\pi\vec{p}) = 4 \sum_{\mu=1}^3 f^2(p_\mu)$,
and the function
$f(p)$ gives the momentum dependent terms in the free fermion propagator,
\beqn
f(p) = {9\over 16} \sin (p) -  {1\over 48} \sin (3p) ~~.
\label{momentum}
\eqn
Results for the free gluon and fermion energy densities are given
in Table~\ref{tab:free}.
\begin{table*}[hbt]
\setlength{\tabcolsep}{1.5pc}
\catcode`?=\active \def?{\kern\digitwidth}

\caption{The energy density of a free gluon ($\epsilon^G$) and a free
fermion ($\epsilon^F$) gas on spatially infinite lattices with temporal
extent $N_\tau$ relative to the corresponding values in the continuum,
$\epsilon^G/T^4= 8\pi^2/15$ and $\epsilon^F/T^4= 7n_f\pi^2/20$.
The last two columns give results for the energy density of four-flavour QCD
in the infinite temperature limit relative to the corresponding
continuum result, $\epsilon_{SB}/T^4 = (\epsilon^G+ \epsilon^F)/T^4$.
}
\vskip 5pt
\label{tab:free}
\begin{center}
\begin{tabular}{|r|r|r|r|r|}
\hline
&\multicolumn{3}{|c|}{ } &~\\
&\multicolumn{3}{|c|}{${\cal O}(a^2)$ improved}& standard \\
\hline
~&~&~&
\multicolumn{2}{|c|}{ }\\
$N_\tau$ &
$\epsilon^G (N_\tau) / \epsilon^G$&
$\epsilon^F (N_\tau) / \epsilon^F$&
\multicolumn{2}{|c|}{
$\epsilon (N_\tau) / \epsilon_{SB}$}
\\
\hline
4 &0.986568 &1.269896&1.191737& 1.694917\\
6 &0.997528 &1.005633&1.003397& 1.659614\\
8 &1.000309 &0.963731&0.973822& 1.424757\\
10&1.000253 &0.986795&0.990507& 1.252452\\
12&1.000150 &0.996509&0.997513& 1.156253\\
\hline
\end{tabular}
\end{center}
\end{table*}
We note that in the gluon sector the deviation from the continuum limit
is below 3\% already on $N_\tau=4$ lattices while the fermion action
still shows about 27\% deviations on this size lattice. Still the
cut-off
dependence is drastically reduced compared to the standard staggered
fermion formulation, which is given for comparison in the last column of
Table~\ref{tab:free}.

\section{QCD equation of state}

We have analyzed the thermodynamics of four-flavour QCD using the
improved action defined in Eq.~\ref{actiong}.
We have performed calculations on $16^3\times 4$ and $16^4$ lattices
with quark masses $ma=0.05$ and 0.1 using the hybrid Monte Carlo
algorithm. The inclusion of the contributions from the three-link terms
to the equations of motion is straightforward. 
At each value of the gauge coupling, $\beta$, typically
1500 trajectories of length $\Delta t=0.6$ have been analyzed on the
finite temperature lattice and 500 trajectories on the zero temperature
lattice.

\begin{figure}[htb]
\begin{minipage}{.51\linewidth}
   \epsfig{
       file=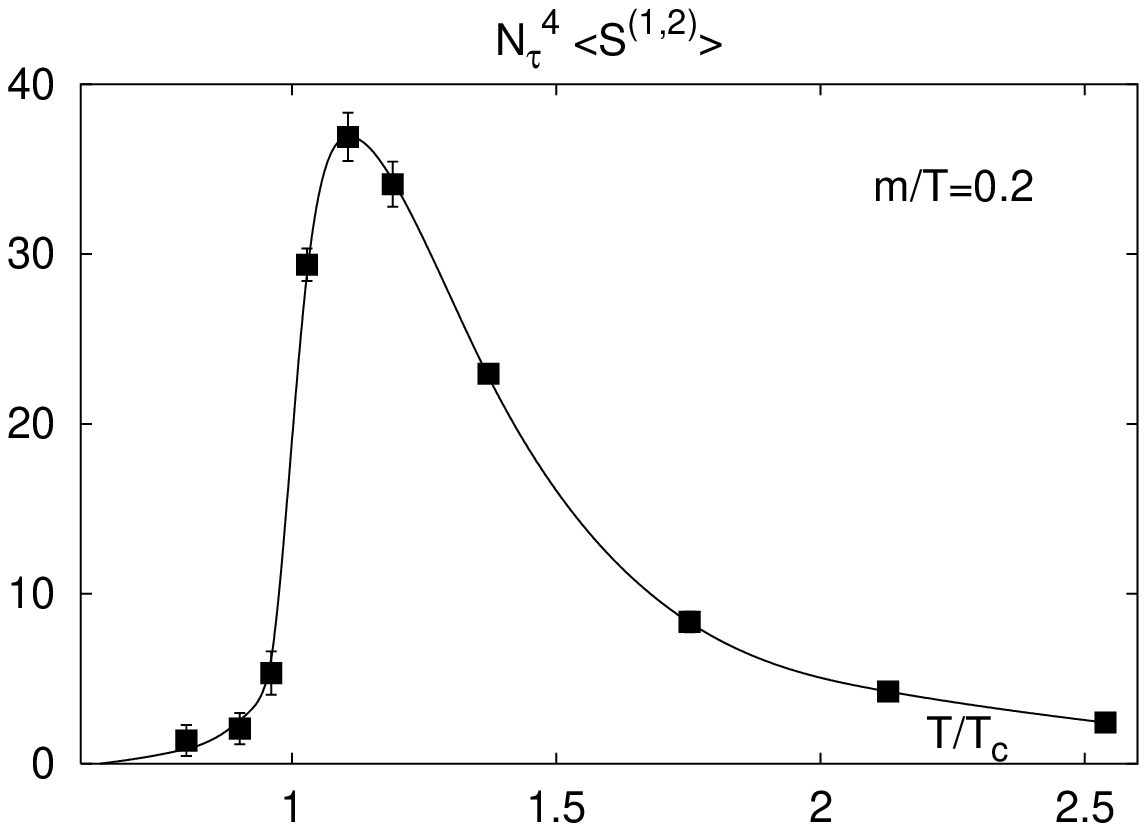,width=75mm,angle=0}
\end{minipage}
\begin{minipage}{.51\linewidth}
   \epsfig{
       file=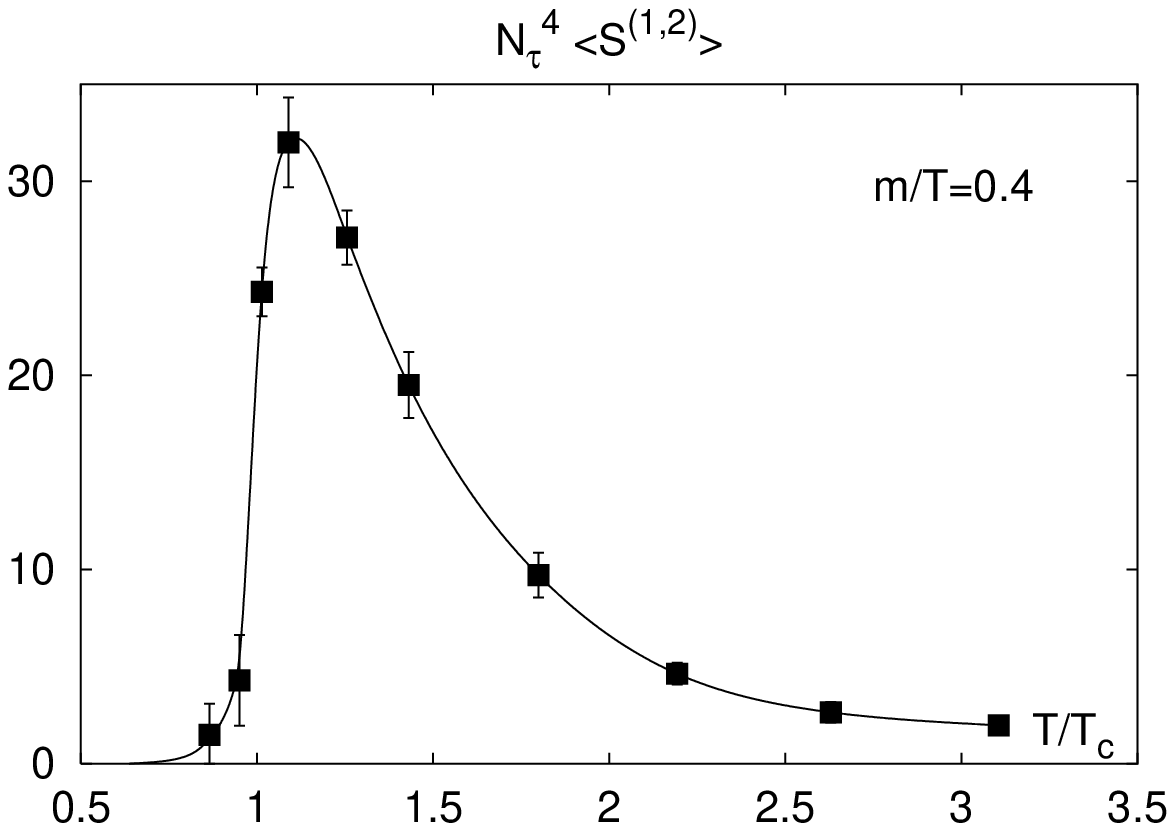,width=75mm,angle=0}
\end{minipage}

\begin{minipage}{.51\linewidth}
   \epsfig{
       file=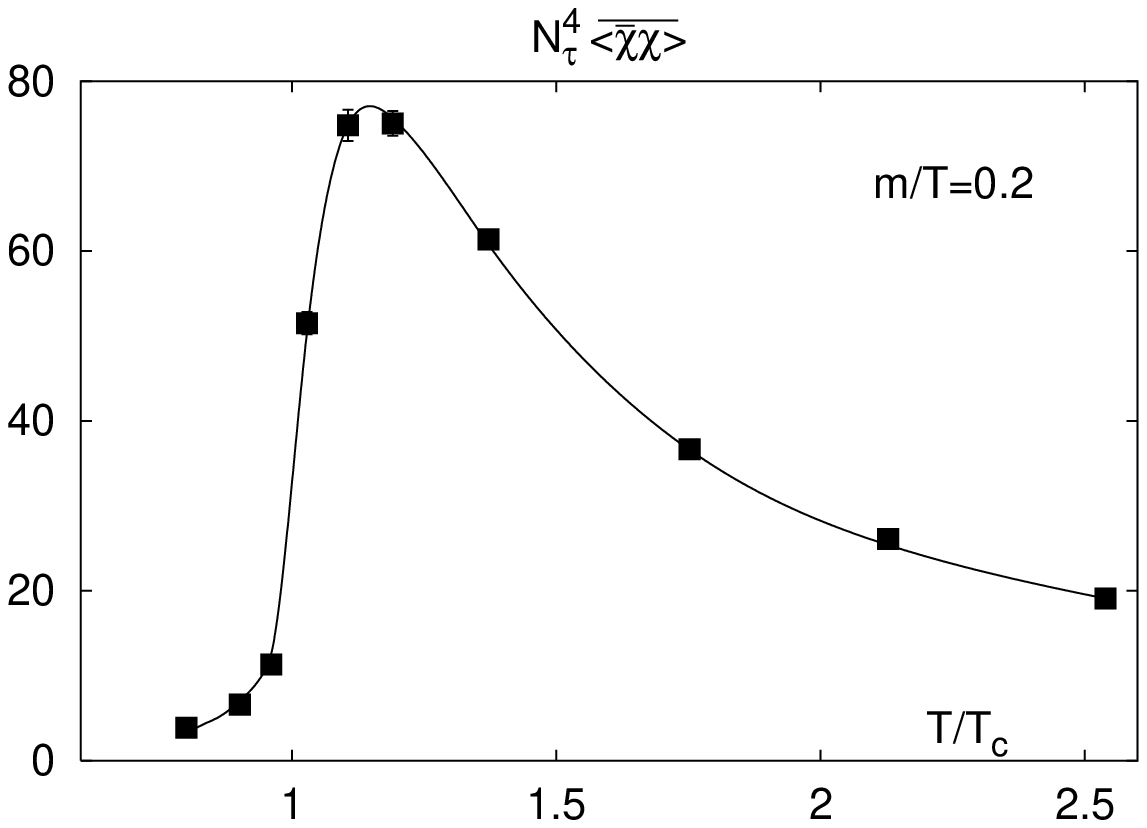,width=75mm,angle=0}
\end{minipage}
\begin{minipage}{.51\linewidth}
   \epsfig{
       file=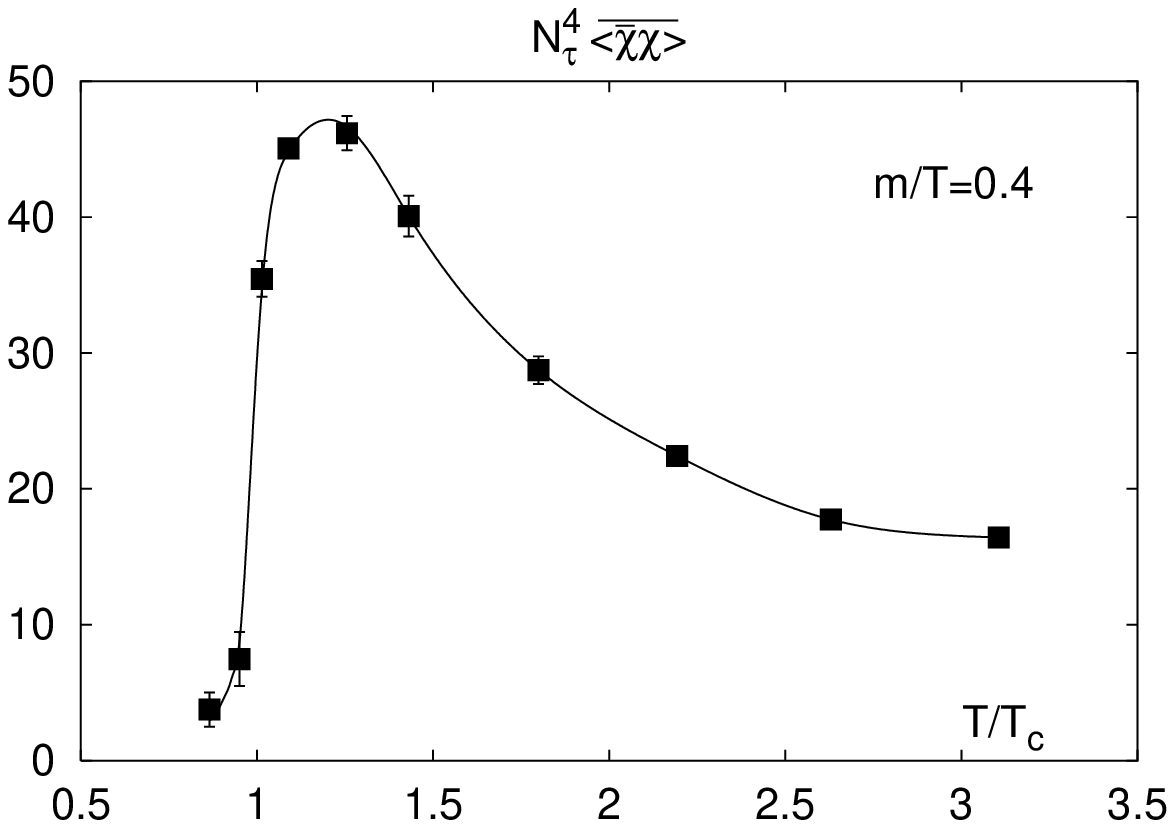,width=75mm,angle=0}
\end{minipage}
\caption{Difference of expectation values of the gluonic action and
chiral condensates
calculated on $16^3\times 4$ and $16^4$ lattices at two values of
the quark mass, $ma=0.05$ and 0.1.}
\label{fig:differences}
\end{figure}

The basic observables entering the calculation of
thermodynamic quantities are the expectation values
for the gauge action, $\langle S^{(1,2)}\rangle$, and the chiral
condensate, $\langle \bar\chi \chi \rangle$, {\it i.e.} the derivatives
of the partition function $Z=\int {\rm d}U {\rm d}\bar\chi {\rm d}\chi
\exp{(-S^I)}$ with respect to the gauge coupling and the quark mass,
respectively. In particular we will
need the differences between expectation values calculated on the finite
temperature lattice ($16^3\times 4$) and the zero temperature lattice
($16^4$),
\begin{eqnarray}
\overline{\langle S^{(1,2)}\rangle} &=&
\langle S^{(1,2)}\rangle_0 -
\langle S^{(1,2)}\rangle_T \nn \\
\overline{\langle \bar\chi \chi \rangle} &=&
\langle \bar\chi \chi \rangle_0 -
\langle \bar\chi \chi \rangle_T ~~,
\label{diff}
\end{eqnarray}
where the normalization of the expectation values in both cases has been
defined per lattice site, {\it i.e.}
$\langle S^{(1,2)}\rangle = -N_\sigma^{-3}N_\tau^{-1}\partial\ln Z
/\partial\beta$ and $\langle \bar\chi \chi \rangle =
N_\sigma^{-3}N_\tau^{-1}\partial\ln Z /\partial ma$.
Our results for these differences are summarized in
Figure~\ref{fig:differences} for both values of the quark mass. The
curves shown in the Figure are spline interpolations, which have been
used
for the subsequent calculation of the pressure ($p$) and energy density
($\epsilon$).

Like in the pure gauge theory the pressure can be obtained from an
integration of the difference of gluonic action densities
$\overline{\langle S^{(1,2)}\rangle}$ \cite{Eng90}
\beqn
{p\over T^4}\Big\vert_{\beta_0}^{\beta}
=~N_\tau^4\int_{\beta_0}^{\beta}
{\rm d}\beta'
\overline{\langle S^{(1,2)}\rangle}~~.
\label{preslat}
\eqn

The calculation of energy density involves in addition also the
differences of the chiral condensates at zero and non-zero temperature.
Furthermore one needs to know the cut-off dependence of the
two bare couplings, $\beta$ and $ma$,
\beqn
\Delta \equiv {\epsilon - 3p \over T^4} = - N_\tau^4\biggl[
R_\beta \overline{\langle S^{(1,2)}\rangle} -
R_m     \overline{\langle \bar\chi \chi \rangle}
\biggr]
\label{delta4}
\eqn
with
\beqn
R_\beta = {{\rm d}\beta \over {\rm d}\ln a}  \quad,\quad
R_m = {{\rm d} ma \over {\rm d}\ln a} ~~.
\label{betafunctions}
\eqn
In the chiral limit the derivative $R_m$
vanishes and $(\epsilon -3 p)$ is again proportional only to the
$\beta$-function, $R_\beta$,
as it is the case in the pure gauge sector. The derivatives of the bare
couplings $\beta$ and $ma$ with respect to the lattice cut-off can be
obtained from the cut-off dependence of two physical
observables. In general we may follow Ref.~\cite{Blu95} and use two
meson masses for this, $m_\pi$ and $m_\rho$. If we parametrize the quark
mass dependence of these masses as
\begin{eqnarray}
a^2 m_\pi^2 &=& h(\beta) ma \quad , \nn \\
a m_\rho &=& f(\beta) + g(\beta) ma \quad ,
\label{parameter}
\end{eqnarray}
we find for the derivatives
\begin{eqnarray}
R_\beta  &=& 
\frac{f}{f'} \frac{1-ma {g \over f} }{ 1 + ma  
( {g' \over g} - {h' \over h} ) {g \over f'} } =
{f \over f'} + {\cal O} (ma) \quad, \nn \\
R_m &=& 2ma \biggl( 1- {h' \over 2 h} R_\beta \biggr) \quad .
\label{derivatives}
\end{eqnarray}
Here the prime denotes derivatives with respect to $\beta$. In the
asymptotic scaling regime the ratios $f/f'$ as well as $h/h'$ will be given
by the leading term in the QCD renormalization group equation, {\it i.e.}
$f/f' =h/h' =-25/4\pi^2$. The second derivative thus becomes
asymptotically
$R_{m} = ma +{\cal O} ((ma)^2)$. We find strong deviations from these
asymptotic relations in the parameter range studied here. This is
immediately evident from the quark mass dependence of the pion mass. To a
first approximation the pion slope turns out to be independent of $\beta$.
This shows that $R_{m} \simeq 2 ma$. In fact, for both quark mass values
and all $\beta$-values studied by us we find this relation to be
satisfied within 15\%.

Rather than using the rho-meson mass to determine $R_\beta$ we have used an
approach which turned out to describe quite well the deviations from asymptotic
scaling in the pure gauge theory. We define an effective coupling,
$\beta_{\rm eff}$, in terms of the expectation value of the gluonic part of
the action,
\beqn
\beta_{\rm eff} = { 12 \over \langle S^{(1,2)} \rangle (\beta) }~~,
\label{betaeff}
\eqn
and use the dependence of $\beta_{\rm eff}$ on $\beta$ to calculate the
derivative $R_\beta$ with the help of the asymptotic two-loop
renormalization group equation. This also fixes the temperature scale,
\beqn
{T \over T_c} = \biggl( {\beta_{\rm eff} \over \beta_c}\biggr)^{-77/625}
\exp{\bigr({4\pi^2/25} (\beta_{\rm eff} - \beta_c)\bigl)} ~~,
\label{scale}
\eqn
We use this approximation for $R_\beta$ also in the calculation of $R_m$
and determine in addition the ratio $h'/h$ from the pion slope.

The critical couplings for both quark masses have been determined from
the
Polyakov-loop susceptibility and the chiral susceptibility. The location
of the peaks in both susceptibilities coincide within errors. From this
we find
\beqn
\beta_c = \cases{
3.57~(3)\quad,\quad ma=0.1 \quad,\cr
3.49~(3)\quad,\quad ma=0.05 \quad, \cr
}
\label{betac}
\eqn
which fixes the temperature scale through Eq.~\ref{scale}.

\begin{figure}[htb]
\vskip -0.7truecm
\vspace{9pt}
\begin{center}
   \epsfig{
       file=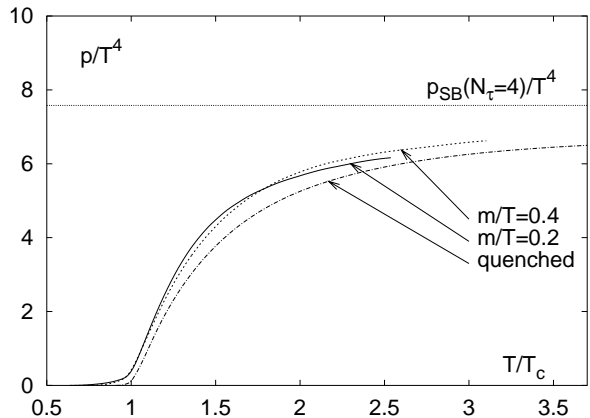,width=85mm}
\end{center}
\vskip -0.7truecm
\caption{Comparison between the pressure of four-flavour QCD on a
$16^3\times 4$ lattice for two values of the quark mass and the pressure of
the pure $SU(3)$ gauge theory. The latter has been rescaled by the appropriate
number of degrees of freedom of four-flavour QCD (29/8) and a factor 1.19
which takes care of the remaining finite cut-off distortion of the ideal gas
limit in our four-flavour simulation (see Table 1).}
\label{fig:pressure4}
\end{figure}

In Figure~\ref{fig:pressure4} we show 
the results of a calculation of the pressure
for both values of the quark mass. As can be seen the result is quite
insensitive to the value of the quark mass. Moreover, we note that the
overall structure of the temperature dependence of the pressure is quite
similar to that of the pure gauge sector \cite{Boy96}.
This also can be seen in Figure~\ref{fig:pressure4}, where we show
the pressure of the $SU(3)$ gauge theory
rescaled by an appropriate ratio of the number
of degrees of freedom so that the high temperature limit coincides with
that of four-flavour QCD.

\begin{figure}[htb]
\begin{minipage}{.50\linewidth}
  \epsfig{
       file=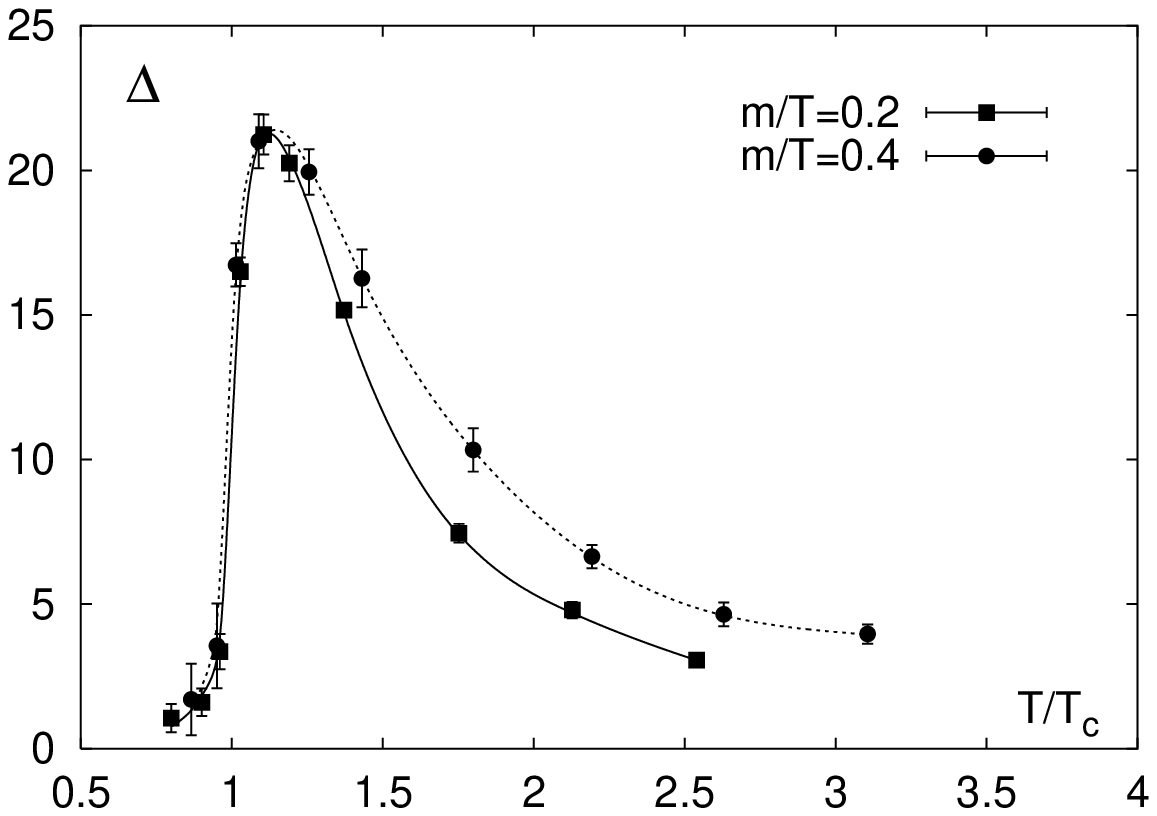,width=78mm,angle=0}
\end{minipage}
\begin{minipage}{.50\linewidth}
  \epsfig{
       file=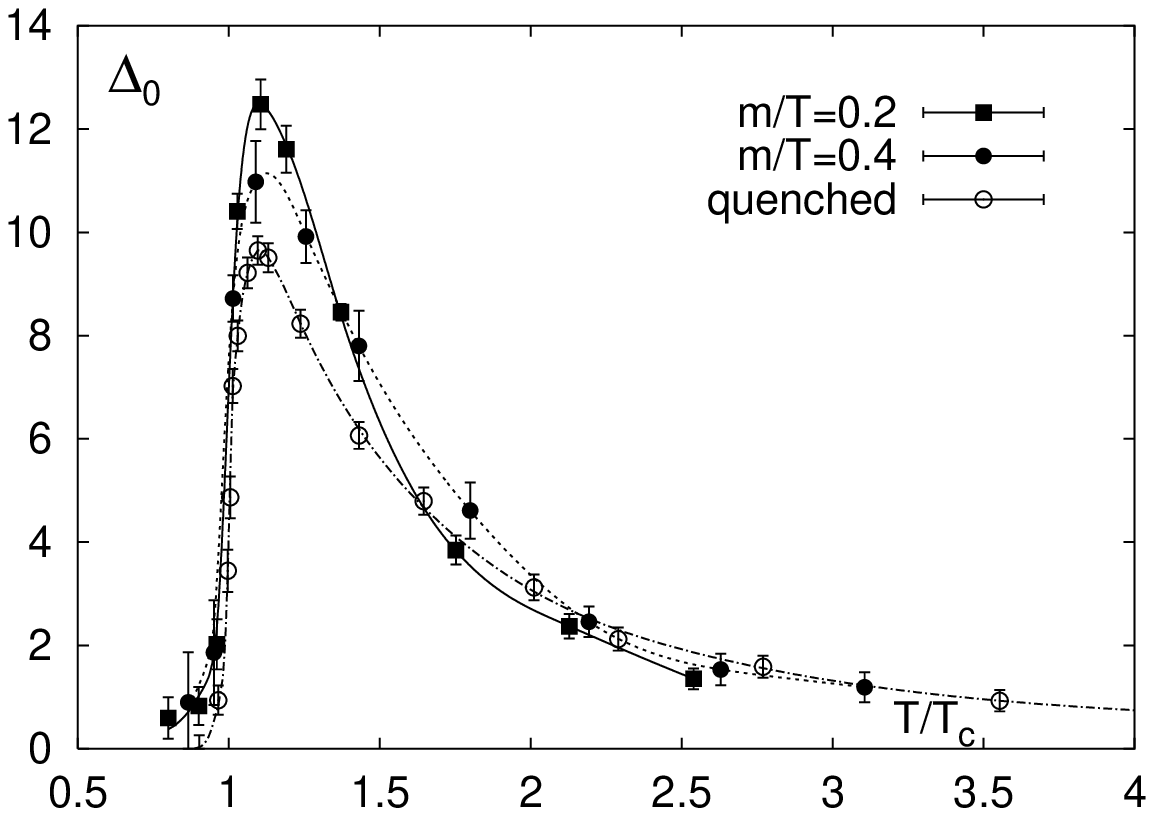,width=78mm,angle=0}
\end{minipage}
\caption{The difference $(\epsilon-3p)/T^4$ calculated according to
Eq.~9 ($\Delta$) as well as Eq.~16 ($\Delta_0$).The lines are spline
interpolations to the data, with the full line for $m/T=0.2$, the
dotted one for $m/T=0.4$ and dash-dotted for the quenched results.}
\label{fig:e3pqcd}
\end{figure}

In Figure~\ref{fig:e3pqcd}a we show the difference between the
energy density and three times the pressure calculated according to
Eq.~\ref{delta4}. As discussed above, the second term in this equation,
{\it i.e.} the term being proportional to the chiral condensates, will
not contribute in the chiral limit. We therefore show in
Figure~\ref{fig:e3pqcd}b also the difference $((\epsilon -3p)/T^4)_0$
defined only in terms of the gluonic part,
\beqn
\Delta_0 = - N_\tau^4 R_\beta
\overline{\langle S^{(1,2)}\rangle}
\label{delta4_0}
\eqn

We note that in the high temperature regime, $T/T_c \gsim 2$, the quark mass
dependence, which is visible in $\Delta$ is completely eliminated
in $\Delta_0$. This suggests that at least in this
temperature
regime $\Delta_0$ can be viewed as a reasonable
extrapolation of
$(\epsilon-3p)/T^4$ to the chiral limit, $ma \rightarrow 0$. Close to $T_c$
this is not that evident. In fact, neither in Figure~\ref{fig:e3pqcd}a nor
in \ref{fig:e3pqcd}b we
see any significant quark mass dependence for $T\lsim 1.2 T_c$.
In the case of $(\epsilon-3p)/T^4$
this is due to the cancellation of two effects in the contribution from the
chiral condensate term. On the one hand the derivative factor
$R_m$ decreases with decreasing quark mass. On
the other hand the difference of chiral condensates,
$\overline{\langle \bar\chi \chi \rangle}$, still increases because the
critical coupling gets shifted towards smaller values (Eq.~\ref{betac}).
As the first effect will ultimately dominate and will force the entire
contribution of the condensate term to vanish in the chiral limit we 
expect that $\Delta_0$ is already a good approximation to the
final result in the chiral limit. Of course, this should eventually be 
checked by performing simulations at smaller quark masses.

Also shown in Figure~\ref{fig:e3pqcd}b is the result for the pure gauge
theory which again has been rescaled with the appropriate number of degrees
of freedom (29/8) as it has been done for the pressure alone. This too gives
support to our extrapolation to the chiral limit and suggests that the
equation of state in four-flavour QCD indeed shows a temperature dependence
which is very similar to that of the pure gauge theory.

\begin{figure}[htb]
\vskip -0.7truecm
\vspace{9pt}
\begin{center}
   \epsfig{
       file=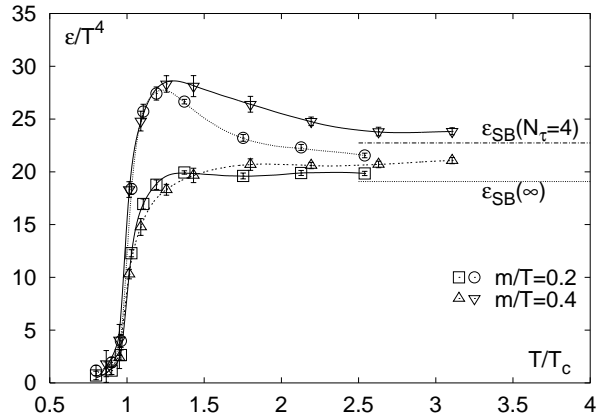,width=85mm}
\end{center}
\vskip -0.7truecm
\caption{Energy density of four-flavour QCD on a $16^3\times 4$ lattice.
The lower set of curves shows an extrapolation to the chiral limit
which has been obtained by ignoring the second term in Eq.~9, {\it i.e.}
from Eq.~16 (see text).}
\label{fig:energy}
\end{figure}

We use the results for the pressure and the difference $\epsilon-3p$ to
extract the energy density. This is shown in Figure~\ref{fig:energy}. The
energy density does come close to the ideal gas limit immediately above
$T_c$. We
do observe an overshooting of the ideal gas limit close to $T_c$ for the
non-zero quark masses considered here. This is a feature not seen
in the pure gauge sector. 
The overshooting does, however, seem to disappear when we determine the
energy density from $\Delta_0$. As discussed above this may be
viewed as an extrapolation to the chiral limit (see also \cite{MILC2}).

\section{The critical temperature}

In the pure gauge sector it has been found that the critical temperature
for the deconfinement transition calculated with the standard Wilson
action in units of the square root
of the string tension, $\sqrt{\sigma}$, deviates by about 10\% from the
continuum extrapolation at $aT_c =1/4$. This cut-off dependence gets
strongly reduced already with tree level improved actions
\cite{Kar96,Bei96b}. To our knowledge the ratio
$T_c/\sqrt{\sigma}$ has previously been determined for the chiral
transition in four-flavour QCD only once using the standard staggered
formulation with a quark mass $m/T=0.08$ at $aT_c = 1/8$ \cite{Alt93}.
This gave $T_c/\sqrt{\sigma} = 0.39~(3)$.

We have analyzed the heavy quark potential on the $16^4$ lattices at the
critical couplings given in Eq.~\ref{betac}. For this purpose we have
generated 300 gauge field configurations separated by 10 trajectories
for the quark mass $ma=0.05$ and 200 configurations for the
larger mass, $ma=0.01$. The potential has then
been extracted using smeared Wilson loops and following the approach
used also in the pure gauge theory \cite{Boy96}.
The potentials have been fitted to an ansatz including a linear and a
Coulomb term, $V_{q\bar{q}}(R) = V_0 + \alpha /R + \hat{\sigma} R$. 
The string tension in units of the cut-off, $\hat{\sigma}\equiv \sigma a^2$
turns out to be significantly larger than in the SU(3) gauge theory
at a comparable value of the cut-off, i.e. $aT_c = 1/4$. The potentials
have been fitted in the range $3 \le R \le 6$. From the fits we obtain
\beqn
T_c/\sqrt{\sigma} = \cases{
0.407~(10) \quad,\quad ma=0.05 \cr
0.430~(~8) \quad,\quad ma=0.1 \cr
}
\label{tcsigma}
\eqn

\section{Conclusions}

We have analyzed the thermodynamics of four-flavour QCD using an
improved fermion action. We find that the ${\cal O} (a^2)$ improvement
of the free fermion action also leads to a strong reduction of the cut-off 
dependence in the high temperature phase. In the ideal gas limit 
deviations from the continuum Stefan-Boltzmann law are still about 20\%.
A further improvement will thus be necessary in order to reduce the
cut-off dependence to only a few percent as it is the case in the pure
gauge sector. In addition one would also like to achieve a reduction
of the flavour symmetry breaking in the staggered action. This does not
seem to be the case for the improved action we have used here
\cite{MILC}.

For the dependence of
the pressure on $T/T_c$ we obtain results which closely follow that of 
the $SU(3)$ gauge theory when rescaled with the number of degrees of freedom.
The same holds for the energy density after an extrapolation to the chiral
limit.
The critical temperature itself drops to $T_c = 0.407(10)\sqrt{\sigma}$ 
at a quark mass $m/T=0.2$.

\end{document}